\documentclass[final]{ias2}

\usepackage{graphicx} 
\usepackage{multirow}
\usepackage{array} 
\usepackage{natbib}

\usepackage{hyperref} 

\begin{document}

\def\d{{\rm d}}\def\e{{\rm e}}\def\Rc{R_{\rm c}}
\def\Gyr{\,{\rm Gyr}}
\def\kms{\,{\rm km\,s}^{-1}}
\def\figref#1{Fig.~\ref{#1}}
%
\font\gkvecten=cmmib10
\font\gkvecseven=cmmib7
\let\boldgrk=\gkvecten
\let\boldgrksc=\gkvecseven

\def\gkthing#1{{\mathchoice%
	{\hbox{{\boldgrk\char#1}}}
	{\hbox{{\boldgrk\char#1}}}
	{\hbox{{\boldgrksc\char#1}}}
	{\hbox{{\boldgrksc\char#1}}}}}

\def\valpha{\gkthing{11}}
\def\vbeta{\gkthing{12}}
\def\vgamma{\gkthing{13}}
\def\vdelta{\gkthing{14}}
\def\vepsilon{\gkthing{15}}
\def\vzeta{\gkthing{16}}
\def\veta{\gkthing{17}}
\def\vtheta{\gkthing{18}}
\def\viotaeta{\gkthing{19}}
\def\vkappa{\gkthing{20}}
\def\vlambda{\gkthing{21}}
\def\vmu{\gkthing{22}}
\def\vnu{\gkthing{23}}
\def\vxi{\gkthing{24}}
\def\vpi{\gkthing{25}}
\def\vrho{\gkthing{26}}
\def\vsigma{\gkthing{27}}
\def\vtau{\gkthing{28}}
\def\vupsilon{\gkthing{29}}
\def\vphi{\gkthing{30}}
\def\vchi{\gkthing{31}}
\def\vpsi{\gkthing{32}}
\def\vomega{\gkthing{33}}
{\newif\ifnotend
\notendtrue
\def\veclist{ABCDEFGHIJKLMNOPQRSTUVWXYZabcdefghijklmnopqrstuvwxyz.}
\def\top#1#2.{#1}
\def\tail#1#2.{#2.}
\loop\expandafter\xdef\csname v\expandafter\top\veclist\endcsname%
{{\noexpand\bf\expandafter\top\veclist}}
\edef\veclist{\expandafter\tail\veclist}
\if\veclist.\notendfalse\fi\ifnotend\repeat}
\def\df{{\sc df}}\def\pdf{{\sc pdf}}\def\aa{{\sc aa}}
\def\LCDM{$\Lambda$CDM}
\def\vmu{\mu}
\def\vlos{v_{\rm los}}

\markboth{Extracting science from surveys of our Galaxy}{James Binney}

\title{Extracting science from surveys of our Galaxy}

\author[sin]{James Binney} 
\email{binney@thphys.ox.ac.uk}
\address[sin]{Rudolph Peierls Centre for Theoretical Physics, University of
Oxford, 1 Keble Road, Oxford OX1 3NP, UK}

\begin{abstract}
Our knowledge of the Galaxy is being revolutionised by a series of
photometric, spectroscopic and astrometric surveys. Already an enormous body
of data is available from completed surveys, and data of ever increasing
quality and richness will accrue at least until the end of this decade.  To
extract science from these surveys we need a class of models that can give
probability density functions in the space of the observables of a survey --
we should not attempt to ``invert'' the data from the space of observables
into the physical space of the Galaxy.  Currently just one class of model has
the required capability, so-called ``torus models''. A pilot application of
torus models to understanding the structure of the Galaxy's thin and thick
discs has already produced two significant results: a major revision of our
best estimate of the Sun's velocity with respect to the Local Standard of
Rest, and a successful prediction of the way in which the vertical velocity
dispersion in the disc varies with distance from the Galactic plane.

\end{abstract}


\pacs{Appropriate pacs here}
\maketitle

\section{Introduction}

When Chandra worked on stellar dynamics, the subject was largely concerned
with understanding the phase-space distribution of near-by stars.
Subsequently the subject's focus shifted to understanding the large-scale
dynamics of globular clusters and external galaxies, for which rich bodies of
observational data started to be available in the 1970s. Recently there has
been renewed interest in the dynamics of our Galaxy as a result of
spectacular improvements in the quantity and quality of data for our Galaxy.

The richness of the available data, and the strong impact that observational
selection effects have upon it, make the task of extracting science from
Galaxy surveys qualitatively different from anything previously attempted in
stellar dynamics. New methods in stellar dynamics, and new methods of data
analysis, will have to be developed if we are to do justice to the data that
will be available by the end of the decade. In this article I describe the
challenge and describe one way of addressing it. 

\section{The current challenge}

We are in the middle of the golden age of surveys of our Galaxy. These
surveys are usefully divided into photometric surveys,
spectroscopic surveys and astrometric surveys.

Important near-infrared photometric surveys include the 2MASS
and DENIS surveys  of the infrared sky, which were completed in the last ten
years, the UKIDSS survey, which is nearing completion and a couple of deeper
surveys that are now getting underway with ESO's VISTA telescope.
The SDSS project, which was completed a few years ago, obtained visual
multi-band photometry of tens of millions of stars

The SDSS project was extended by the SEGUE project, and together these
surveys obtained low-resolution spectra of several hundred thousand stars.
The RAVE survey, which is nearing completion, will provide spectra at
resolution $R=7\,500$ for $\sim500\,000$ stars. The LAMOST telescope, which
is currently being commissioned, will take optical spectra of huge numbers of
halo stars.  The APOGEE project, which aims to gather 100\,000 near-infrared
spectra in the northern hemisphere at $R=30\,000$, will soon be taking data.
Meanwhile the HERMES project will obtain a similar number of near-infrared
spectra in the southern hemisphere, and a large spectral survey with the VLT
is to be undertaken by ESO.

Astrometric astronomy was revolutionised by the European Space Agency's
Hipparcos mission, which published a catalogue of $\sim100\,000$ parallaxes
in 1997. Hipparcos established an all-sky reference frame that was tied to
quasars. The US Naval Observatory used this reference frame to re-reduce a
large body of terrestrial observations, leading to the UCAC3 catalogue, which
gives proper motions for $10^8$ stars. The Pan-Starrs survey is beginning to
image much of the sky to magnitude $V\sim24$ on a regular basis. It will
discover enormous numbers of variable stars and measure parallaxes and proper
motions for all the objects it detects. In 2013 ESA will launch Gaia, the
follow-on to Hipparcos and the first satellite to conduct an astrometric
survey of the sky -- Hipparcos had an input catalogue while Gaia will itself
identify objects. Gaia will obtain astrometry of unrivalled precision down to
magnitude $V\sim20$ and spectra for objects brighter than $V\sim17$. In all
the Gaia Catalogue, which should be published around 2020, will contain
astrometry for $\sim10^9$ stars and stellar parameters and line-of-sight
velocities for $\sim10^8$ stars.

\section{How predictive is $\Lambda$CDM?}

The Cold Dark Matter model of cosmology, latterly with the modification that
vacuum energy contributes an appreciable cosmological constant, has been a
spectacular success in accounting for the pattern of fluctuations in the
cosmic background radiation and the clustering of galaxies. In light of these
successes it is often stated that the goal of surveys of the Milky Way is to
``test \LCDM''.  

I think this point of view over-estimates the predictive power of the \LCDM\
model, which is after all a theory of the invisible. Its successes are based
on the ease with which the clustering of collisionless matter can be computed
when it dominates the gravitational field. In particular, before the era of
decoupling, the photon/baryon fluid was nearly homogeneous, and until the era
of the first stars ($z\simeq15$?), the baryons were no more strongly
clustered than the dark matter (DM) so the gravitational field was everywhere
dominated by the DM distribution.  Consequently, the distribution of
DM can be quite reliably computed from the linear regime up to the era
of the first stars.

From that time on the theory becomes enormously more intractable because
crucial events were occurring in regions dominated by the gravitational field
of the baryons, and that field depends on the tremendously complex physics of
baryons: strong interactions (nucleosynthesis), weak interactions (supernova
blast waves), and electromagnetic interactions (stellar winds and photo-heating
to name but two key processes) all have a major impact of the distribution of
matter and thus on the driving gravitational field.

For more than a decade a relatively small group of courageous theorists have
been endeavouring to include key aspects of baryon physics in simulations of
cosmic evolution. Although they are extremely complex and challenging, these
simulations are not truly ``ab-initio'' in the sense that they take the
well-understood physics of stars as read and merely try to follow how stars
form, and how energy that is released by them impacts the surrounding
interstellar medium. Truly ab-initio simulations require a dynamic range that
is so large that it will probably never be possible, and we do not know what
dynamic range has to be attained to make reliable simulations of the type
that are currently being undertaken. Until that dynamic range has been
attained, we do not know what predictions \LCDM\ makes for the structures of
galaxies. 

The job of surveys of the Milky Way is to discover what is out there, and we
should approach this job in a open-minded spirit.

\section{Discrete models are not enough}

N-body simulations have been of crucial importance for the development of our
understanding of stellar dynamics. They have, moreover, been key to the
development of the \LCDM\ model, and irrespective of any reservations one may
have about the extent to which model Galaxies extracted from cosmological
simulations are based on irrefutable physics, these models have had a big
impact on our thinking, and they will undoubtedly play a significant role in
the future.  Nevertheless, I will now argue that we need other types of
models too.

The major issue with N-body models is the difficulty of fitting them to data.
This problem has several aspects

\begin{itemize} 
\item A high-quality model is computationally expensive to produce, and the
connection between its structure and the initial conditions and
phenomenological model from which it started is unclear. Consequently, it is
hard to know how the initial conditions or input sub-grid physics should be
modified to obtain a model that provides a better fit to given data. This
objection has recently been weakened by the development of ``made-to-measure''
(M2M) modelling \citep{SyerTremaine,deLorenzi07}. M2M modelling allows one to
steer an N-body model towards a better fit to given observational data, and it
has already produced the best current model of the inner Galaxy
\citep{Bissantz}. 

\item The precise arrangement of the model's particles is of no interest (a
short time later every particle will be somewhere else); the model's content
is the underlying probability density function (pdf) of which the current
particle distribution is just a discrete realisation. The precise
distribution of material in the Galaxy is likewise a discrete realisation of
an underlying pdf. If we had the pdf of the model, we could evaluate the
likelihood of the Galaxy given the model. But we don't have this pdf, and
asking whether two discrete realisations are consistent with a common pdf is
hard. The obvious way to accomplish this task is to estimate the pdf of one
of the two realisations by binning the particles, but we shall see in \S6
below
that this procedure is problematic in the case of the Galaxy.

\item Our location within the Galaxy leads to our seeing right down the
luminosity function in the immediate vicinity of the Sun, but being able to
detect only the most luminous stars far away in the disc or at the Galactic
Centre. Crucial information is carried by low-luminosity objects seen only
locally (for example old white dwarfs encode the history of star formation),
and luminous stars are obviously important tracers of the Galaxy's
large-scale structure. So models need to encompass the entire range of
luminosities. However, as \cite{Brown+} pointed out, it is not clear how an
N-body model can fulfil this criterion: if each particle represents a star,
most particles will have to represent low-luminosity stars, so they will be
invisible anywhere but near the Sun, while if particles represent cohorts of
stars, a particle that lies near the Sun will give rise to many
low-luminosity stars, all with identical kinematics. Again the essential
point is that we really need the underlying pdf, not a discrete realisation
of it.

\end{itemize}

\noindent I conclude that N-body models are not suited to the extraction of
science from survey data because they lack flexibility and do not provide the
required pdfs.

\section{Steady-state models are crucial}

The Galaxy is not in a steady state, most obviously because it contains a bar
near the Centre, and spiral structure within the disc, and more subtly
because the SDSS survey revealed that the stellar halo is largely comprised
of streams and still-enigmatic ``clouds'' like the Hercules-Aquila Cloud
\citep{Belokurov}. None the less, modellers of the Galaxy have no option but
to start by constructing steady-state models. 

The reason for this necessity
is that DM makes a major contribution to the Galaxy's gravitational field
-- our best guess is that the Sun lies close to the radius at which
baryon-domination at small radii gives way to DM-domination at large radii.
At present, we can detect DM only through its contribution to the
gravitational field, which we map through the influence it has on objects
that we can see -- on Galactic scales this amounts to studying the dynamics
of gas and stars. In principle {\it any\/} phase-space distribution of stars
is consistent with {\it any\/} gravitational field. It is only when we insist
on some sort of statistical equilibrium that the distribution of stars
imposes constraints on the gravitational field, and thus on the combined
density of stars, gas and DM. For example, the assumption of statistical
equilibrium enables us to rule out a weak gravitational field, because in
that field the observed distribution of stars would expand systematically,
while a very strong gravitational field can be excluded because it
would cause the observed distribution of stars to contract systematically.

The argument just given is rather crude in that is does not engage with the
more subtle devices for determining the gravitational field, such as the use
of hydrodynamical simulations of the flow of gas \citep{EnglmaierG,Sellwood+}, or
stellar streams \citep{Johnston+,EyreB4}.  However, I believe these techniques also
rest on the assumption that the Galaxy is broadly in statistical equilibrium.

Actually, even if we could observe DM directly, and hence determine the
Galaxy's gravitational field without recourse to dynamics, it would still
make sense to seek an equilibrium model of the Galaxy first. Then comparing
the predictions of this model with the data we would identify features that
signalled departures from equilibrium, and we could seek to model these
features by perturbing our equilibrium model. In this connection it is
salutary to recall the extent to which the language of physics, and thus our
understanding of phenomena, has been moulded by perturbation theory:
dispersion relations, photons, phonons, Feynman diagrams, orbital elements,
mean-motion resonances, etc., are all concepts, introduced by perturbation
theory, that loom large in our understanding of how the world works. Thus they
are both useful mathematical abstractions and essential tools for
understanding. Historically, perturbation theory has been rather little used
in galactic dynamics, and we are all the poorer in understanding for it.

\section{We need to fit models to data in data space}

We like to conceive of the Galaxy as an object that lives either in
three-dimensional ``real'' space, or better in six-dimensional phase space.
Actually, below I shall argue that even the most basic Galaxy model inhabits
a space of at least ten dimensions, but for the moment let's be conservative
and imagine that it inhabits phase space, where we can readily write down the
equations that govern the evolution in time of the pdf of a system of
mutually gravitating particles. 

Unfortunately, we do not directly measure the natural coordinates of this
space. For example, Gaia will measure two angular coordinates $(\alpha,\delta)$ and a
parallax $\varpi$ instead of a distance. For many stars it will measure
the line of sight velocity $\vlos$ and for all objects it will measure two
components of the proper motion $\vmu$. So for many stars Gaia will measure
six coordinates that form rather an odd set from the perspective of physics.
In particular, the star's physical location and two of its components of
velocity depend on the measured parallax, which for a distant star may be
measured to be negative. Clearly a negative parallax is unphysical, but it
does carry information: it tells us that the star is more distant than the
distance that corresponds to the uncertainty in the parallax. Our modelling
strategy must be such that we can make good use of stars with negative
measured parallaxes

Because the inverse of a negative parallax cannot be interpreted as a
distance, a strategy that is {\it not\/} going to work is to infer the star's
phase-space coordinates from the data; we cannot carry the star from the
space of the data into the space of the model, namely phase space. We {\it must\/}
proceed in the opposite direction, projecting the model into the space of the
observables. When we carry the model into the space of observables, negative
parallaxes are in no way problematic -- indeed a catalogue in which negative
parallaxes did not occur is what would be problematic.

Even for stars with safely positive parallaxes, there is a huge advantage in
carrying the model into the space of observables because in this space the
errors are likely to be largely uncorrelated and even Gaussian. Indeed, they
will be the result of a large number of statistically independent chance
events: the displacement of the measured photocentre from its true location
is determined by photon-counting statistics, distortions of the CCD's grid,
etc. The pdf in phase space into which a Gaussian pdf in
$(\alpha,\delta,\varpi,\vlos,\mu_\alpha,\mu_\delta)$ space would map would be
strongly non-Gaussian and predict large correlations between errors in
distance
and the components of tangential velocity, for example. With such a complex
pdf for the errors, we have no hope of achieving a satisfactory understanding
of how errors give rise to uncertainties in the parameters of our models.

Given that we must carry our models into the space of observables, let us
examine more carefully what coordinates that space really has. In addition to
$(\alpha,\delta,\varpi,\vlos,\mu_\alpha,\mu_\delta)$ we will invariably
measure an apparent magnitude $m$, a colour such as $V-I$. The spectrum from
which $\vlos$ was extracted will also have yielded the star's surface gravity
$\log g$ and measures of metallicity, such as [Fe/H] and possibly
[$\alpha$/Fe]. If a medium- or high-resolution spectrum is available, there
will be measurements of the abundances of many individual chemical elements,
such as C, O, Mg, Ti, Eu, etc. Hence, the dimensionality of the space of
observables $(\alpha,\delta,\varpi,\vlos,\mu_\alpha,\mu_\delta,m,V-I,\log
g,[\hbox{Fe/H]},\ldots)$ will normally be as high as 10, and it may be significantly
higher. Our modelling strategy must be designed to cope with such large
dimensions.

The high dimensionality of the space of data makes any analysis that involves
binning the data very unattractive. To see this, imagine establishing a
Cartesian grid in data space, the cell spacing in most dimensions being of
order half the typical observational uncertainty of the corresponding
observable. If we applied this criterion to $l$ and $b$, we would obtain an
absurdly fine grid on the sky. Therefore, in $l$ and $b$ we take the
separation between bins to be comparable to the changes in angle over which
we expect the distribution of stars to change significantly -- this might be
as large as $10^\circ$ in $l$ and a degree or so in $b$ at low latitudes and
larger increments near the poles.  Having established our grid, we assign
stars to cells and thus obtain the density of stars in each bin. The Poisson
noise in our density estimates decreases with the number of stars
assigned to each cell, which increases with the adopted bin sizes and
decreases with the number of quantities actually observed. Hence binning is
most attractive for the sparsest data set, namely measurements of
$(\alpha,\delta,m,\mu_\alpha,\mu_\delta)$. However, this is already a five-dimensional space. For a
survey of a third of the sky, it might be possible to get by with 100 bins on
the sky. A few tens of bins would be required for the apparent magnitudes,
and for each component of $\vmu$ ten bins might suffice. Thus a minimal grid
will have $>10^5$ bins, so even with the simplest conceivable data, binning
first becomes advantageous when the number of stars in the catalogue exceeds
a million. In the case of Gaia, the list of observables must be expanded to
include at least $\varpi$, and the number of bins required to do justice to
the proper-motion data must be increased by a factor of at least 10, implying
a grid with $>10^6$ bins.  In reality one would want to include colour data,
and some line-of-sight velocities, and the number of bins would be pushed up
to $\sim10^9$, essentially the number of stars in the catalogue. Thus we
should not rely on binning the data.

\section{We must use multiple lines of evidence}

If a star is near enough, no measure of distance can trump a parallax. But
usually the majority of the objects measured in a survey lie far away, and
then a parallax measurement carries less information. For these objects
spectrophotometric distances are likely to be important. In many cases a
small parallax will imply that the star is distant, and from its apparent
magnitude it will be evident that it is a giant. This information can inform
the choice of template star used in the analysis of the star's spectrum and
thus the determination of its values of $\vlos$, $\log g$ and [Fe/H], as well
as its distance. Thus the pdf of the distance upon which we
finally settle will depend on  several sources of information and a great
deal of modelling.

Several of the quantities we measure are connected by well-understood
physics. For example, the theory of stellar evolution constrains stars to a
small subset of $(T_{\rm eff},\log g,\hbox{[Fe/H]})$ space. The principles of
Bayesian inference give us a framework for using such constraints to reduce
the uncertainties in stellar parameters \citep[e.g.][]{BurnettB}. Ideally we
would extract stellar parameters in parallel with fitting a Galaxy model to
the data, since the stellar parameters depend on the distance, just as the
parallax and tangential velocities do. This scheme appears to be hard to
implement in practice, so in the near future stellar parameters extracted
from spectrophotometric data in isolation will play a large role in Galaxy
modelling. We should not lose sight, however, of the long-term goal of
expanding the ``observables'' to include the calibrated spectrum and fitting
the model to the data in this expanded space.

\section{Torus models}

I believe only one class of models fulfils the requirements we have identified above,
namely, of being steady-state models that yield a pdf in the space of
observables, and  can be used as a basis for studying non-equilibrium
features by means of perturbation theory. This is the class of ``torus
models'', which I will now describe.

Torus modelling is essentially a variant of Schwarzschild modelling
\citep{SchwarzI}, which
has been for more than a decade the standard tool with which to extract
dynamical information from observations of external galaxies (especially in
connection with searches for massive central black holes \citep{Gebhardt03}). To build a
Schwarzschild model one guesses the form of the system's gravitational
potential and then integrates a representative ``library'' of orbits in the
chosen potential. Finally one seeks non-negative weights for each orbit such
that the weighted sum of the contributions of each orbit to the observables
is consistent with the data. If we can find no satisfactory solution with the
current gravitational potential $\Phi$, then we adjust $\Phi$ and try again.
 
Torus modelling involves replacing orbits,
which are time series of phase-space points, by
orbital tori, which are analytic expressions for the three-dimensional
continuum of phase-space points that are accessible to a star on a given
orbit. Whereas an orbit is labelled by its initial conditions, a torus is
labelled by its three action integrals $J_i$; whereas position on an orbit is
determined by the time $t$ elapsed since the initial condition, position on a
torus is determined by the values of three angle variables $\theta_i$, one
canonically conjugate to each action $J_i$. For a
summary of how orbital tori are constructed and references to the papers in
which torus dynamics was developed, see \cite{McMB08}.

Replacing numerically integrated orbits with orbital tori brings the following
advantages:

\begin{itemize}

\item{} The density of stars in phase space is simply related to the sampling
density of tori in action space and the weights assigned to those tori. This
is because tori have prescribed actions and the six-dimensional phase-space
volume occupied by orbits with actions in $\d^3\vJ$ is
$\tau=(2\pi)^3\d^3\vJ$. Knowledge of the phase-space density of orbits allows
one to convert between orbital weights and the value taken by the \df\ on an
orbit. It proves advantageous to choose an analytic representation of the
\df\ $f(\vJ)$ and to derive the weights of individual tori from $f$. The
weights are varied by adjusting parameters that appear in $f$ \citep{B10}.

\item{} There is a clean and unambiguous procedure for sampling orbit space.
The choice of initial conditions from which to integrate orbits for a library
is less straightforward because the same orbit can be started from many
initial conditions, and when the initial conditions are systematically
advanced through six-dimensional phase space, the resulting orbits are likely
at some point to cease exploring a new region of orbit space and start
resampling a part of orbit space that is already represented in the library.
On account of this effect, it is hard to relate the weight of an orbit to the
value taken by the \df\ on it \citep[but see][for how this can be
done]{Haefner,Thomas05}.

\item{} There is a simple relationship between the distribution of stars in
action space and the observable structure and kinematics of the model -- as
explained in \S4.6 of \cite{BT08}, the
observable properties of a model change in a readily understood way when
stars are moved within action space.  The simple relationship between the
observables and the distribution of stars in action space enables us to infer
from the observables the approximate form of the \df\ $f(\vJ)$, which is nothing but the
density of stars in action space.

\item{} From a torus one can readily find the velocities that a star on a
given orbit can have when it reaches a given spatial point $\vx$. By contrast
a finite time series of an orbit is unlikely to exactly reach $\vx$, and
searching for the time at which the series comes closest to $\vx$ is
laborious. Moreover, several velocities are usually possible at a given
location, and a representative point of closest approach must be found for
each possible velocity.

\item{}
An orbital torus is represented by of order 100 numbers while a
numerically-integrated orbit is represented either by some thousands of
six-dimensional phase-space locations, or by a similar number of occupation
probabilities within a phase-space grid.

\item{} The numbers that characterise a torus are smooth functions of the
actions $\vJ$. Consequently tori for actions that lie between the points of
any action-space grid can be constructed by interpolation on the grid.
Interpolation between time series is not practicable.

\item{} Schwarzschild and torus models are zeroth-order, time-independent
models which differ from real galaxies by suppressing time-dependent
structure, such as ripples around early-type galaxies
\citep{MalinC,QuinnP,SchweizerF}, and spiral structure or warps in disc
galaxies. Since the starting point for perturbation theory is action-angle
variables \cite[e.g.][]{Kalnajs}, in the case of a torus model one is well
placed to add time-dependent structure as a perturbation.
\cite{Kaasalainen95a} showed that classical perturbation theory works
extremely well when applied to torus models because the integrable
Hamiltonian that one is perturbing is typically much closer to the true
Hamiltonian than in classical applications of perturbation theory
\citep{GerhardS,DehnenG,Weinberg}, in which the unperturbed Hamiltonian
arises from a potential that is separable (it is generally either spherical
or plane-parallel).

\end{itemize}

The impact of shot noise on the model is usually minimised if all tori have
the same weight, and this will be the case if the density of used tori in
action space samples the \df. We endeavour to ensure that this condition is
met, at least to a good approximation \citep{BMcM11}.

\section{The interface between the thin and thick discs}

A considerable body of evidents now points to the disc of our Galaxy being a
superposition of a ``thick disc'' made up of old stars on orbits that are
moderately eccentric and inclined to the Galactic plane, and a thin disc of
stars with higher ratios of Fe to O and Mg abundances that are on less strongly
eccentric or inclined orbits. The details of the superposition are, however, murky because
stars from both  populations are found near the Sun and even at the same
velocities. 

\cite{B10} proposed assigning a simple \df\ to each population. For the thick disc
he proposed
 \begin{equation}
f_{\rm thk}(J_r,J_z,L_z)=f_{\sigma_r}(J_r,L_z)f_{\sigma_z}(J_z),
\label{thickDF}
\end{equation} 
 where
 $f_{\sigma_z}$ is defined by  
 \begin{equation}
f_{\sigma_z}(J_z)\equiv{\e^{-\Omega_z J_z/\sigma_z^2}
\over2\pi\int_0^\infty\d J_z\,\e^{-\Omega_z J_z/\sigma_z^2}}.
\label{basicvert}
\end{equation}
 Here $\Omega_z(\vJ)$ is the fundamental frequency of vertical oscillations,
$\sigma_z$ is a constant with the dimensions of velocity, and the denominator ensures that $f_{\sigma_z}$ satisfies the normalisation
condition
 \begin{equation}
\int\d z\,\d v_zf_{\sigma_z}=1\quad\Leftrightarrow\quad 
\int\d J_z\,f_{\sigma_z}={1\over2\pi}.
\label{normf}
\end{equation}
 Similarly, $f_{\sigma_r}$ is defined by
\begin{equation}
f_{\sigma_r}(J_r,L_z)\equiv{\Omega\Sigma\over\pi\sigma_r^2\kappa}\bigg|_{\Rc}
[1+\tanh(L_z/L_0)]\e^{-\kappa J_r/\sigma_r^2}.
\label{planeDF}
\end{equation}
 Here $\Rc(L_z)$ is the radius of the circular orbit with angular momentum
$L_z$,
\begin{equation}
\Sigma(L_z)=\Sigma_0\e^{(R_0-\Rc)/R_\d}
\end{equation} 
 is the thick disc's surface density, and $\Omega(L_z)$ and $\kappa(L_z)$
are the circular and the radial epicycle frequencies there. With these
choices the disc's surface density is approximately exponential in $R$ with
scale length $R_\d$. We take $L_0\ll R_0v_{\rm c}(R_0)$ so the term in square
brackets serves to eliminate retrograde stars.

The scale heights of the discs of external galaxies seem to be approximately
independent of radius, and this finding suggests that the vertical velocity
dispersion in these discs is roughly proportional to $\e^{-R/2R_\d}$. We
enforce similar behaviour on the radial and vertical dispersions within the
thick disc by taking both $\sigma_r$ and $\sigma_z$ to be functions of $L_z$:
 \begin{eqnarray}
\sigma_r(L_z)&=&\sigma_{r0}\,\e^{q(R_0-\Rc)/R_\d}\nonumber\\
\sigma_z(L_z)&=&\sigma_{z0}\,\e^{q(R_0-\Rc)/R_\d}.
\end{eqnarray}
 The only important parameters of $f_{\rm thk}$ are $\sigma_{r0}$ and
$\sigma_{z0}$.

For the \df\ of the thin disc, \cite{B10} proposed a superposition of
``pseudo-isothermal'' \df s like (\ref{thickDF})
 \begin{equation}
f_{\rm thn}(J_r,J_z,L_z)={\int_0^{\tau_{\rm m}}\d\tau\,\e^{\tau/t_0}
f_{\sigma_r}(J_r,L_z)f_{\sigma_z}(J_z)
\over t_0(\e^{\tau_{\rm m}/t_0}-1)}.
\label{thinDF}
\end{equation}
 The idea here is that in the thin disc star formation has continued
throughout
the lifetime of the Galaxy at a rate that has declined exponentially with
time constant $t_0\simeq3\Gyr$. Throughout the lifetime of a cohort of
coeval stars,  scattering of its stars by non-axisymmetric
fluctuations in the gravitational potential has increased its velocity
dispersions roughly as a power law of age, $\sigma\propto\tau^\beta$, so in equation (\ref{thinDF}) the
dispersion parameters are given by
 \begin{eqnarray}
\label{sigofLtau}
\sigma_r(L_z,\tau)&=&\sigma_{r0}\left({\tau+\tau_1\over\tau_{\rm
m}+\tau_1}\right)^\beta\e^{q(R_0-\Rc)/R_\d}\nonumber\\
\sigma_z(L_z,\tau)
&=&\sigma_{z0}\left({\tau+\tau_1\over\tau_{\rm m}+\tau_1}\right)^\beta
\e^{q(R_0-\Rc)/R_\d},
\end{eqnarray}
 where $\sigma_{r0}$ and $\sigma_{z0}$ are constants.

\begin{figure}
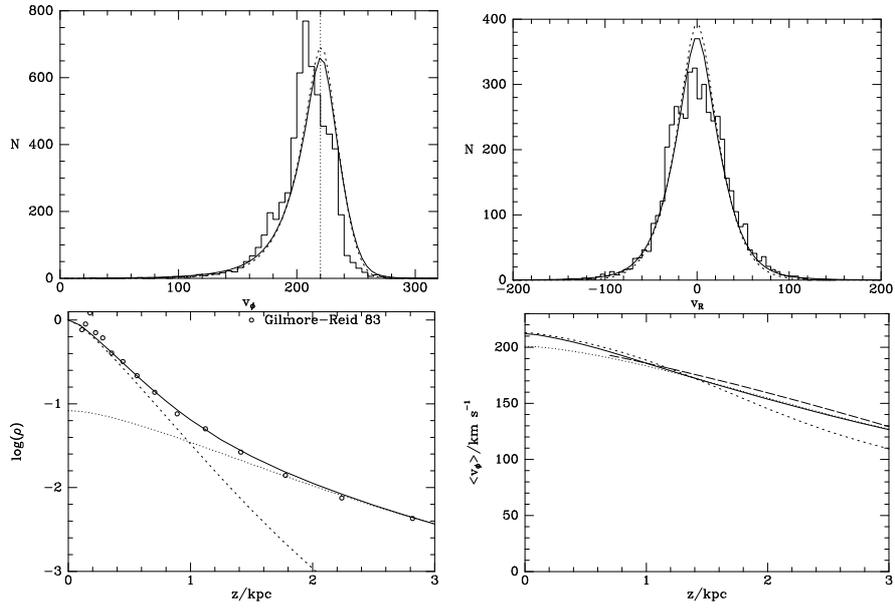

\centerline{\includegraphics[width=.45\hsize]{VphiTT4.ps}\quad
\includegraphics[width=.45\hsize]{VRTT4.ps}}
\centerline{\includegraphics[width=.45\hsize]{rhozTT4.ps}\quad
\includegraphics[width=.45\hsize]{vphizTT4.ps}}
 \caption{Structure at the solar radius predicted by the standard \df\
(eqs.~\ref{thinDF} and \ref{thickDF}).  Full curves are for the entire disc
while dashed curves show the contribution of the thin disc.  The upper panels
show velocity distributions for $z=0$ marginalised over the other velocity
components. The vertical dotted line marks the local circular speed.  In the
lower panels dotted curves show the contributions of the thick disc. In the
bottom-right panel the long-dashed line is the empirical fitting function of
Ivezic et al.\ (2008).  The values of the \df's parameters are given in Table 1.}
\label{fig:thinthick}
\end{figure}

\figref{fig:thinthick} shows prediction of the complete \df\ for the
structure of the solar neighbourhood when $\Phi_z(z)$ is taken to be the
potential above the Sun in Model II of \S2.7 in \cite{BT08}; this model is
disc-dominated. Full curves are for the whole disc and dashed curves show
the contribution of the thin disc. The upper panels show for stars seen in
the plane the distributions in $v_\phi$ and $v_R$ after integrating over the
other two velocity components; the smooth curves are the predictions of the
model, while the histograms show the distributions observed in the
Copenhagen--Geneva survey \citep{HolmbergNA}. The lower left panel shows that
the model provides a good fit to the vertical density profile reported by
\cite{GilmoreR}, who discovered the thick disc. The lower-right panel
shows the prediction of the model for how the mean-streaming speed should
decrease with height above the plane. 

\subsection{The Sun's azimuthal velocity}

The major weakness of the fits between model and data displayed in
\figref{fig:thinthick} is the displacement in the top left panel of the model
curve to higher values of $v_\phi$ than the histogram. It proves impossible
to rectify this problem by changing the \df. 

The histogram is here plotted as a function of $v_\phi$, the azimuthal
velocity in the Galactic rest frame, whereas the data derive from
measurements of velocities relative to the Sun. Consequently, the histogram
can be moved to the right if we increase our estimate of the Sun's velocity
relative to the local standard of rest. From Hipparcos data \cite{DehnenB98}
determined this to be $5.25\pm0.62\kms$ by plotting the mean motion relative
to the Sun of different groups of stars, versus the square of the group's
random velocity.  A naive reading of Stromberg's equation \citep[eq.~(4.228)
in][]{BT08} predicts that this
plot will be linear \S4.8.2(a) of \cite{BT08}, and the Hipparcos
data showed that it was apart from the two points with the lowest random
velocities, which must be affected by inadequate phase mixing of freshly
formed stars. What \cite{DehnenB98} overlooked is that by grouping stars by
colour they introduced a tendency for the groups with blue colours and low
velocity dispersions to contain metal-poor stars, and vice versa for the
groups of red stars. On account of the metallicity gradient in the disc, the
guiding-centres of metal-poor stars tend to be at larger radii than those of
metal-rich stars, so when the are visiting the solar neighbourhood,
metal-poor stars tend to have larger $v_\phi$ than metal-rich stars. That is,
the metallicity gradient modifies the relation between colour and
$v_\phi$ that \cite{DehnenB98} used, and thus influenced their value of
$V_\odot$. When \cite{SchoenrichBD} fitted an updated form of the Hipparcos
data to a model of the chemodynamical evolution of the disc, they found
$V_\odot=12\pm2\kms$, $11\sigma$ larger than the old value. When the smooth
curve in the top left panel of \figref{fig:thinthick} is shifted to the right
in accordance with the new value of $V_\odot$, it agrees with the data quite
nicely, and in fact \cite{B10} proposed $V_\odot=11\kms$ for just this reason.

\begin{figure}
\centerline{\includegraphics[width=.7\hsize]{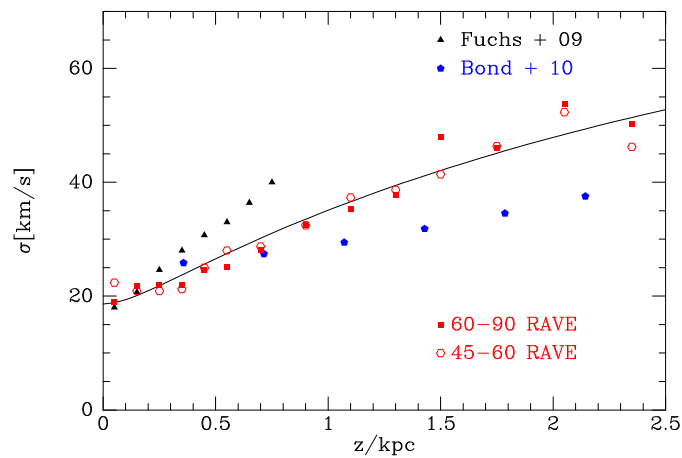}}
\caption{The variaton of $\sigma_z$ with distance from the Galactic plane.
The triangles and filled circles show conflicting determinations from the
SDSS survey. The full curve is the model of \cite{B10}. The red squares and
circles show the values subsequently determined from the RAVE survey by
\cite{Burnett10}.}\label{fig:Wz}
\end{figure}
\subsection{Dispersion in $W$ above the Sun}

A second nice example of the value of models of the disc that have
distribution functions analytic in $\vJ$ is shown in \figref{fig:Wz}. The
lowest and highest series of points show estimates of $\sigma_z(z)$ extracted
from the SDSS by \cite{Bond+} and \cite{Fuchs+}, respectively. These two
estimates are clearly incompatible with each other.  The smooth curve shows
the run of $\sigma_z$ with $z$ in a model of the disc that \cite{B10} obtained by
fitting the data for $\rho(z)$ plotted in \figref{fig:thinthick}. \cite{B10} shows
that this model also provides a good fit to the distribution of $v_z$ of the
GCS stars, although this fit was not used to determine the model's
parameters. He was unable to find a model that was consistent with both these
data and either the data sets derived from the SDSS. The red squares
and circles in \figref{fig:Wz} show estimates of $\sigma_z(z)$ that were
subsequently extracted from the RAVE survey \citep{Steinmetz} by
\cite{Burnett10}: the squares use stars at $b>60^\circ$, while the circles
use stars at $45^\circ<b<60^\circ$. The agreement between the two sets of
points is excellent, which inspires confidence in the spectrophotometric
distances to the RAVE stars that \cite{Burnett10} used. The new data points also agree
very well with the model of \cite{B10}. Thus we have a second example of data that
proved impossible to fit using an analytic \df\ turning out to be
themselves flawed, and the predictions of the \df\ being vindicated.

\section{Conclusion}

We are in the middle of a golden age for surveys of our Galaxy. We already
have an enormous body of data, and data of ever more spectacular quality will
continue to become available until at least the end of this decade.
Extracting from these data an understanding of the current structure and the
history of the Milky Way is going to be a formidable challenge. The sheer
quantity and heterogeneity of the data make the task hard. The task is
made harder still by our peculiar position within the Galaxy itself, as a
consequence of which all surveys contain strong selection effects. Worse
still,
one of the crucial observables, parallax, can be easily scattered by measuring
errors into negative values, which have no physical interpretation.

In light of these difficulties our strategy must be to build models from
which we can compute the pdf of the observed quantities and to use this pdf
to compute the likelihood of the data given the model. From this likelihood
we can compute the pdf of the model's parameters.

While N-body models are a key tool for developing our understanding of both
galactic dynamics and cosmology, they appear to be unsuited to the
interpretation of Galactic surveys because they do not deliver pdfs and they
are hard to steer towards a model that is consistent with the data. A more
promising way forward is offered by torus models, which are analogous to
Schwarzschild models except that orbits (time-series of phase-space points)
are replaced by orbital tori (three-dimensional subspaces of phase space). In
principle the weights of individual tori can be independently assigned, as
are the weights of orbits in a conventional Schwarzschild model, but a better
strategy is to derive the weights from an analytic \df\ $f(\vJ)$, and to vary
these weights by varying a relatively small number of parameters in $f$.

The effectiveness of this strategy is demonstrated by using it to construct
\df s for the thin and thick discs and adjusting its parameters to fit data
from the Hipparcos, RAVE and SDSS surveys. This exercise led to the
uncovering of a subtle error in the determination of the Sun's velocity with
respect to the local standard of rest, and led to the successful prediction
of the variation of $\sigma_z$ with distance $z$ above the Galactic plane.

While this early work shows great promise, it does not conform to the
methodology advocated here in that it fits the model to ``data'', such as the
number of stars in a range of values of $v_\phi$, that are in reality the
result of carrying the data from the space of observables to the physical
space to yield $\overline{v}_\phi$, $\rho(z)$ and $\sigma_z(z)$, for example.
We are currently testing our ability to recover the \df\ by fitting to the
data in the space of measured quantities by fitting \df s to pseudo-data
constructed from similar \df s.  First results are extremely encouraging in
that they show that with only 10\,000 stars the parameters of the \df\ can be
recovered to good accuracy even when the stars are drawn from the (broad)
general luminosity function, and the only data available are sky positions,
proper motions and apparent magnitudes. Complementing the data with
measurements of parallax or line-of-sight velocity reduces the already small
errors and especially correlations between the errors in the model's
parameters.

There is a great deal of work to do in this field, but the  outlook is
exciting.

\end{document}